\documentclass[12pt]{iopart}

\usepackage{setstack}
\usepackage{iopams}

\usepackage{color}
\usepackage{times}
\usepackage{graphicx}
\usepackage{epstopdf}
\usepackage{float}
\usepackage{threeparttable}
\usepackage{fullpage}

\begin{document}

\title{Prospects for early localization of gravitational-wave signals from compact binary coalescences with advanced detectors.}

\author{Alessandro Manzotti$^{12}$, Alexander Dietz$^2$}

\ead{manzotti.alessandro@gmail.com}
\address{$^1$Dipartimento di Fisica, Universita' degli Studi di Parma, Italy}
\address{$^2$Department of Physics and Astronomy, The University of Mississippi, University, MS 38677-1848, USA}

\date{\today}

\begin{abstract}
A leading candidate source of detectable gravitational waves is the inspiral and merger of pairs of stellar-mass compact objects. The advanced LIGO and advanced Virgo detectors will allow scientists to detect inspiral signals from more massive systems and at earlier times in the detector band, than with first generation detectors. The signal from a coalescence of two neutron stars is expected to  stay in the sensitive band of advanced detectors for several minutes, thus allowing detection \textit{before} the final coalescence of the system.

In this work, the prospects of detecting inspiral signals prior to coalescence, and the possibility to derive a suitable sky area for source locations are investigated. As a large fraction of the signal is accumulated in the last $\sim$10 seconds prior to coalescence, bandwidth and timing accuracy are largely accrued in the very last moments prior to coalescence. We use Monte Carlo techniques to estimate the accuracy of sky localization through networks of ground-based interferometers such as aLIGO and aVirgo. With the addition of the Japanese KAGRA detector, 
it is shown that the detection and triangulation before coalescence may be feasible. 
\end{abstract}
\pacs{04.30.Tv, 04.80.Nn, 98.70.Rz}

\maketitle

\section{Introduction}

In the near future advanced gravitational wave (GW) detectors, like advanced LIGO (aLIGO)\cite{advLIGO}, advanced Virgo (aVirgo)\cite{advVirgo} and the Japanese KAGRA\footnote[1]{formerly known as Large scale Cryogenic Gravitational wave Telescope (LCGT)}~\cite{0264-9381-27-8-084004} will be operative, allowing us to regularly observe gravitational wave signals. 
Nowadays there is a growing awareness that coincident observations of GW signals and electromagnetic (EM) counterparts will play a crucial role in the future of GW astronomy allowing us to distinguish between the cases of environmental rather than astrophysical origins of events in the data with a relatively low signal. 
Such observations will also be very important for astrophysics and cosmology.
As an example, information regarding the supernova engine could be obtained by GW signals from core-collapse supernovae, GW signals coincident with gamma-ray bursts could help in determining the nature of the progenitor, and signals from a binary coalescence allow us to measure independently the distance and redshift to the source \cite{Nissanke:2009kt}.
This possibility will help us to determine the Hubble parameter to within a few percent \cite{DelPozzo:2011yh,Dalal:2006qt}. 


Because the major part of electromagnetic observatories are intrinsically directional, a possible approach to perform multi-messenger observations is to localize the position of the source from the gravitational wave observations as quickly and accurately as possible to use electromagnetic observatories to follow up the event. 
Then the first goal of GW astronomy would be to extract the sky location from the gravitational wave signal with enough accuracy. However, conversely to electromagnetic telescope, gravitational wave detectors are sensitive to most of the sky and it is not possible to obtain directional information of a short duration GW source from a single GW detector. Indeed the location of a GW signal is primarily obtained using triangulation methods based on the observed time delays of the signal at different detectors and, for this reason, a network of two or more detectors is crucial for multi-messenger astronomy. 
With the new era of network with three or more advanced detectors a better localization capability will be surely achieved.
For example, a challenging project to follow up gravitational wave sources from initial GW detectors using electromagnetic telescopes is already underway \cite{Kanner:2008zh,Branchesi:2011mi}. 
Localizing the GW sources with a good accuracy is not enough to perform multi-messenger observations, it must be done as rapidly as possible. Fortunately, the improvement of sensitivity in advanced detectors allow some GW signals from inspiral binaries to stay in the aLIGO sensitive band for several minutes, giving the possibility of an early detection.

Regarding EM detection, a great number of electromagnetic observatories with a a field of view (FOV) of around ten~square degree like Pan Starrs, Palomar Transient Factory, LOFAR, SkyMapper and many others (see \cite{LVC:2011ys} for an overview) are being specifically designed for transient phenomena.
Obviously a good estimate of localization accuracy in the case of early detected GW signals will help us to understand the feasibility of the future follow-up searches.

In this work we examine the ability of gravitational wave detectors to detect and localize transient signals \textit{before} the coalescence. 
We perform a Monte Carlo simulation of sources distributed uniformly in space for two different mass scenarios with the LIGO-Virgo (LV) and LIGO-Virgo-KAGRA (LVK) detector network. We compute the probabilities  to detect a merger prior to coalescence, and the associated sky area. 

The layout of this paper is as follows. In section~\ref{sec:sources} the source population is described  for this investigation, and section~\ref{sec:detection} describes how we calculate the SNR values of the simulated sources and how to obtain the sky area. 
Section~\ref{sec:results} will summarize the results which are discussed in section~\ref{sec:conclusion}.

\section{Source Population}
\label{sec:sources}

\subsection{Waveforms} 

In this work we will consider non spinning inspiral waveforms. Only the inspiral part of the waveform is used, which ends at the innermost stable circular orbit (ISCO) with frequency $f_{isco}=\frac{c^3}{6\sqrt6 \pi G M}$. Here $c$ is the speed of light, $G$ is the gravitational constant and $M$ is the total mass of the merger system.
The frequency evolution of the waveform can be written as~\cite{Brown:2004vh}

\begin{equation}
f(t)=\frac{c^3}{8 \pi  G M \Theta (t)^{3/8}} \;,
\label{eq:ft}
\end{equation}

\noindent
where

\begin{equation}
\label{eq:theta}
\Theta(t)=\frac{c^3 t \eta }{5 G M} \;,
\end{equation}

\noindent
and $\eta=\frac{m_1\,m_2}{(m_1+m_2)^2}$ is the symmetric mass ratio. 
Eqs.~(\ref{eq:ft})-(\ref{eq:theta}) can be used to find the frequency of the system as a function of time. 
A measured signal $h(t)$ can be written as

\begin{equation}
h(t)=F_{+}h_{+}(t)+F_{\times}h_{\times}(t) \;.
\end{equation}

\noindent
where $h_{+}(t)$ and $h_{\times}(t)$ are the + and  $\times$ polarization of the signal from the source, and  $F_{+}$ and $F_{\times}$ are the antenna response factors, which depend on the relative orientation between the location of the binary to the antenna~\cite{Anderson:2000yy, Brown:2004vh}. 
As these factors depend on the location of the source relative to the detector, they are time dependent. 
Even for a low-mass binary system and a detector threshold of 10~Hz, the time the signal stays in the sensitive band of a detector is well below an hour. As most of the signal-to-noise ratio (SNR) is accumulated close to coalescence, the effects of Earth rotation can be neglected.

\subsection{Signal population}
\label{sec:pop}

We consider two populations of sources, one consisting of double neutron stars (NS--NS) with masses of 1.4$M_\odot$ each, and one consisting of a neutron star with a black hole (NS--BH), with masses of 1.4$M_\odot$ and 10$M_\odot$, respectively.
For the advanced detectors we assume two lower cutoff frequencies, one at $f_{\mathrm{min}}=30$~Hz and a more optimistic at  $f_{\mathrm{min}}=10$~Hz.
Systems consisting of two black holes are not considered in this work, as the signal would coalesce at much smaller frequencies and the time spend in the sensitive detection band would be very short. Furthermore, it is unlikely that such systems would produce bright electromagnetic counterparts.

The merger rate is assumed to follow the stellar birth rate in spiral galaxies, which can be determined by blue galaxy luminosities~\cite{Kopparapu:2007ib}. However, there is reason to believe that the blue luminosity alone is not a good tracer of the merger rate, since a significant delay between the formation of the compact objects and their coalescence is expected.
Therefore we assume a substantial fraction of merger to take place in old elliptical galaxies \cite{Kennicutt:2009jg, O'Shaughnessy:2009ft}.
Since the range to which advanced detectors are sensitive is hundreds of Mpc, it is safe to assume a spatially uniform distribution of merger sources.

\section{Signal detection}
\label{sec:detection}

\subsection{Single detector SNR calculation}
\label{sec:snr}
 
\begin{figure}[t!]
\begin{center}
\includegraphics[scale=0.4]{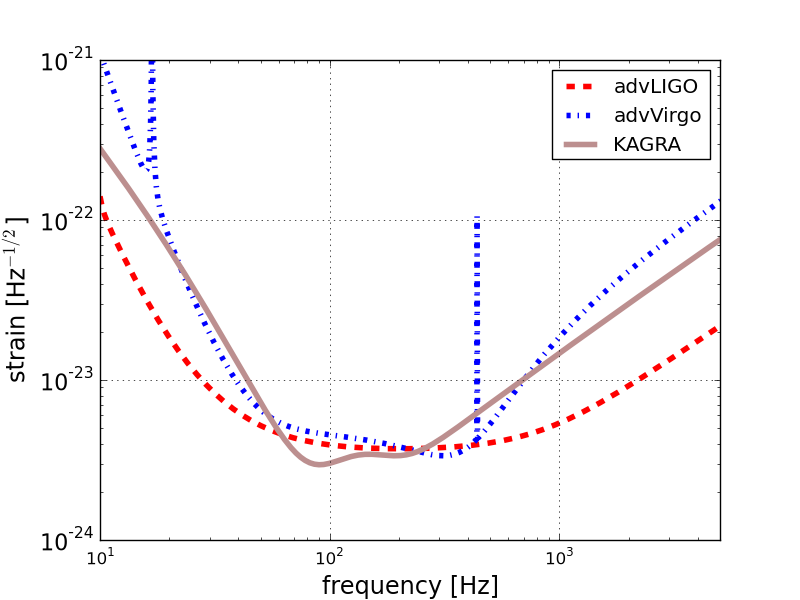}
\caption{$\sqrt{S(f)}$ Power spectrum for Advanced LIGO, Advanced Virgo and KAGRA.}
\label{powers}
\end{center}
\end{figure}

To find a signal in the data, we use standard matched filter algorithms~\cite{Allen:2005fk, LIGO:2010cfa}. The SNR of the signal is

\begin{equation}
\rho=\sqrt{4\int_{f_{min}}^{f_{isco}}\frac{h^*(f)h(f)}{S_{h}(f)}df}\;,
\label{eq:snr0}
\end{equation}

\noindent
where $h^*(f)$ is the complex conjugate of $h(f)$ and $S_{h}(f)$ is the power spectral density (PSD) function of the detector. The anticipated noise curves for advanced LIGO \cite{T0900288} and advanced Virgo~\cite{aVirgoCurve} are plotted in Fig. \ref{powers}. 
As the Japanese KAmioka GRAvitational wave detector (KAGRA)\ might also be operational later in this decade, it is important to include this detector in this investigation. 
The power spectral density of KAGRA is available in text-form~\cite{kagra}, which can be sufficiently fitted by the analytic expression

\begin{eqnarray}
\sqrt{S_{h}(f)} &=& 6.499\cdot 10^{-25} \times \bigg[ \\ \nonumber
&+&  9.72\cdot 10^{-9}  \exp{\left( -1.43-9.88x-0.23x^2 \right)} \\ \nonumber
&+& 1.17 \exp{\left( 0.14-3.10x-0.26x^2 \right)} \\  \nonumber
&+& 1.70 \exp{\left( 0.14+1.09x-0.013x^2 \right)} \\ \nonumber 
&+& 1.25 \exp{\left( 0.071+2.83x-4.91x^2 \right)} \bigg]\\ \nonumber 
\end{eqnarray}

\noindent
where $x = \log(f/100 \mathrm{Hz})$. 

The advanced Virgo noise curve can be parametrized as 

\begin{eqnarray}
\sqrt{S_{h}(f)} &=& 1.259\cdot 10^{-24} \times \bigg[ \\ \nonumber
&+& 0.07 \exp{\left( -0.142-1.437x+0.407x^2 \right)} \\ \nonumber
&+& 3.10 \exp{\left( -0.466-1.043x-0.548x^2 \right)} \\ \nonumber
&+& 0.40 \exp{\left( -0.304+2.896x-0.293x^2 \right)} \\ \nonumber
&+& 0.09\exp{\left( 1.466+3.722x-0.984x^2 \right)} \bigg]\\ \nonumber
\end{eqnarray}

\noindent
where $x = \log(f/300 \mathrm{Hz})$. 

The central idea of this work is to estimate the accumulated SNR of a GW source before the coalescence of the system at the ISCO frequency.
This is done by changing the upper limit on the integral in Eq. (\ref{eq:snr0})

\begin{equation}
\rho(f_{max})=\sqrt{4\int_{f_{low}}^{f_{max}}\frac{h^*(f)h(f)}{S_{h}(f)}df}\;,
\label{Snr}
\end{equation}

\noindent
and then using Eq. (\ref{eq:ft}) to determine the maximum frequency $f_{\mathrm{max}}$ as a function of time. 
The SNR can be expressed as~\cite{VirgoBaseline}

\begin{equation}
\rho=1.56 \times 10^{-19} \left( \frac{M}{M_{\odot}} \right)^{5/6}  \left( \frac{\mathrm{Mpc}}{R} \right) F_{geo} \int_{f_{\mathrm{low}}}^{f_{\mathrm{max}}}\frac{f^{-7/3}}{S_{h}(f)}df\;,
\end{equation}

\noindent
where 

\begin{equation}
F_{\mathrm{geo}}=\sqrt{F_{+}^{2}(1-\cos{\iota}^{2})^{2}/4+F^{2}_{\times}\cos{\iota}^{2}}
\label{eq:fgeo}
\end{equation}

\noindent
represents the response of an antenna in the direction of the source, $\iota$ being the angle between the axis of orbital angular momentum of the binary and the line of sight to Earth.

\begin{figure}[h!]
\begin{center}
\includegraphics[scale=0.3]{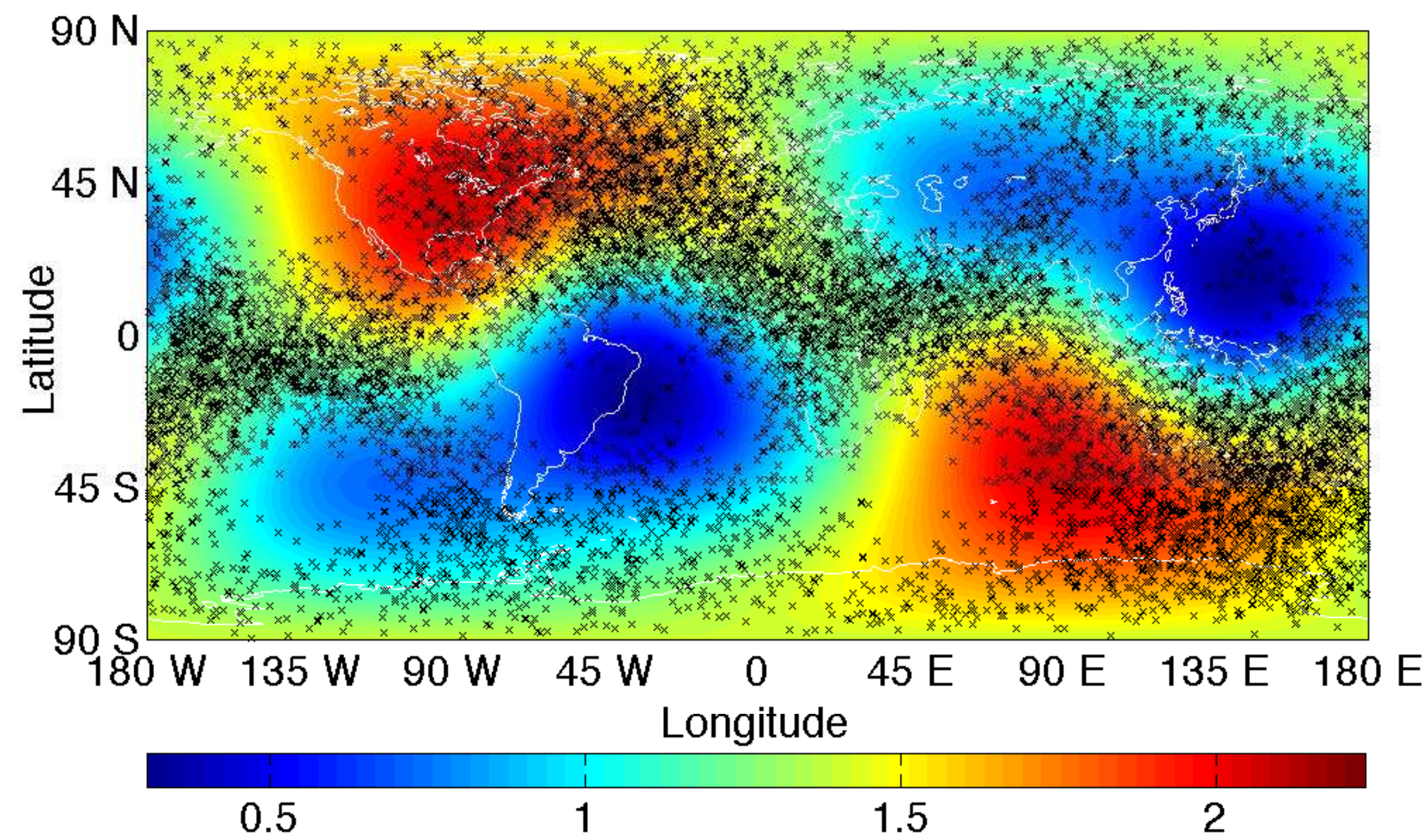}
\caption{Network antenna power pattern over the sky $ \sum_{k}(F_{+,k}+F_{\times,k})$ for a three-detector network, consisting of the two aLIGO detectors and aVirgo. The network antenna power pattern describes the network relative sensitivity in different directions. Crosses represent 10000 detected sources}
\label{fig:response-network}
\end{center}
\end{figure}

\begin{figure}[h!]
\begin{center}
\includegraphics[scale=0.3]{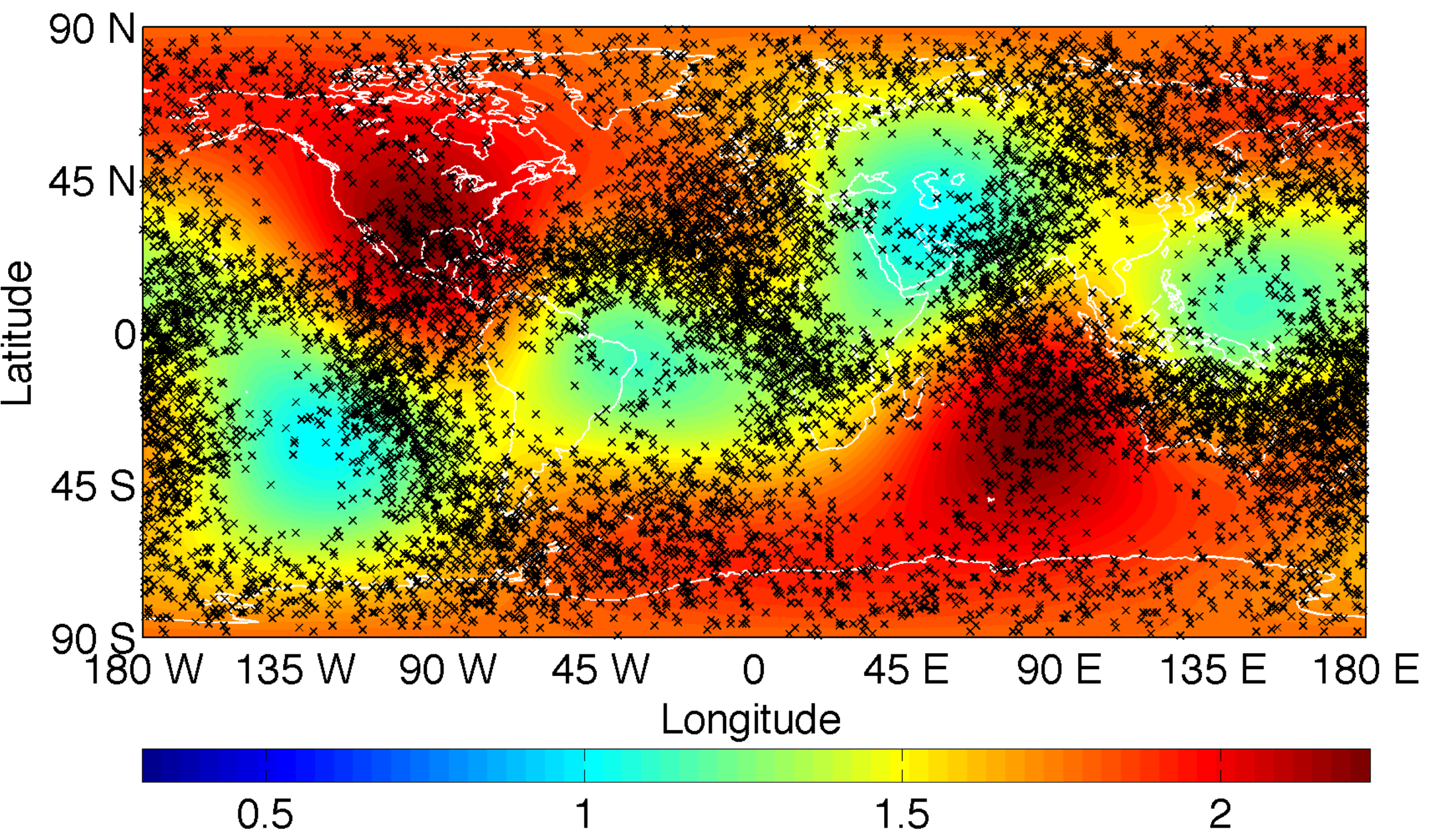}
\caption{Network antenna power pattern $\sum_{k}(F_{+,k}+F_{\times,k})$ over the sky for a four-detector network, consisting of the two aLIGO detectors, aVirgo and the Japanese KAGRA detector. The network antenna power pattern describes the network relative sensitivity in different directions. Crosses represent 10000  detected sources.}
\label{fig:response-network4}
\end{center}
\end{figure}

\subsection{Detector Network}
\label{sec:network}

We now examine two different detector networks.
The first network is made up of advanced LIGO and advanced Virgo (LV), while the second network includes the Japanese KAGRA detector (LVK).
The expected PSD for advanced LIGO, advanced Virgo and KAGRA is used to calculate the estimated SNR of a source. 

A signal is considered to be \textit{detected} by the network, if the recovered SNR exceeds a threshold of 7 in each detector of the network~\cite{Aylott:2009ya}. 
Although a reasonable detection of a signal can be made with two detectors alone, useful triangulation can only be done with three or more detector sites in the network.

The response of the three-detector network to sources in the sky is shown in Fig.~\ref{fig:response-network}. 
This plot shows the response of a given position above earth to the LV network.
Each cross in this Figure represents a source in the sky that is being detected by the simulation, i.e. for which the recovered SNR is at least 7 in each detector. 
Fig.~\ref{fig:response-network4} shows the same result for the LVK network. The colorscale shows the network antenna power pattern $ \sum_{k}(F_{+,k}+F_{\times,k})$~\cite{Schutz:2011tw}, where $k$ goes over each detector. This is a generalization of the antenna power pattern for a network of detectors.
It describes the relative sensitivity of the network in different directions.

\begin{figure}[t!]
\begin{center}
\includegraphics[scale=0.45]{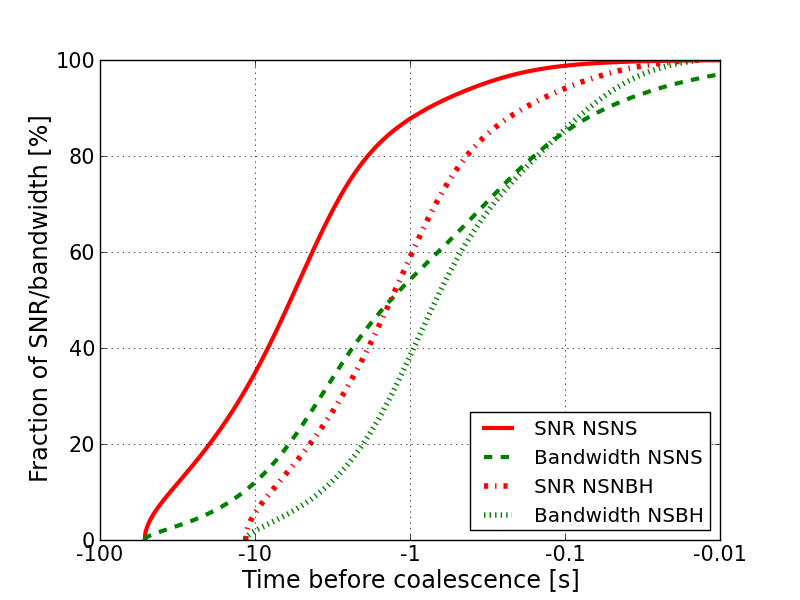}
\caption{SNR $\rho$ and signal bandwidth $\sigma_{f }$ as a function of time prior to merger. Although a large fraction of the SNR is accumulated well before the coalescence, the bandwidth increases significantly at later times. The signal of a NS--NS binary enters the sensitive band at 52 seconds prior to coalescence (for a 30~Hz lower frequency cutoff), while it is 12 seconds for a NS--BH signal.}
\label{fig:snr_band}
\end{center}
\end{figure}

\subsection{Localization of the source}
\label{sec:loc}

Given a list of detected sources, it is now possible to estimate the accuracy of localization for each source. 
For the purposes of this work, we assume the localization is achieved entirely by triangulation between the detectors although the individual amplitudes of the signal provide additional information at the $\sim$10\% level~\cite{DietzSkyLoc}. 

The accuracy of the observed time of coalescence $\sigma_t$  depends on the SNR of the signal and the bandwidth $\sigma_f$ of the detector's response. An approximation of this relationship can be found in Ref.~\cite{Fairhurst:2009tc}:

\begin{equation} \label{accuracy}
\sigma_{t}=\frac{1}{2 \pi \rho \sigma_{f}}\;.
\end{equation}

The timing accuracy is inversely proportional to the SNR $\rho$ and the effective bandwidth $\sigma_{f}$ of the source.

The approximations used to obtain Eq.~(\ref{accuracy}) break down at low SNR, where second order effects become important. For a single detector SNR threshold of 7, this approximation differs from the exact calculation by $\sim$10\%, which is within the accuracy of this work (see right panel of Fig.~1 in Ref.~\cite{Fairhurst:2009tc}). 

\subsubsection{Localization with three detectors}
\label{sec:loc3}

\begin{figure}[t!]
\begin{center}
\includegraphics[scale=0.3]{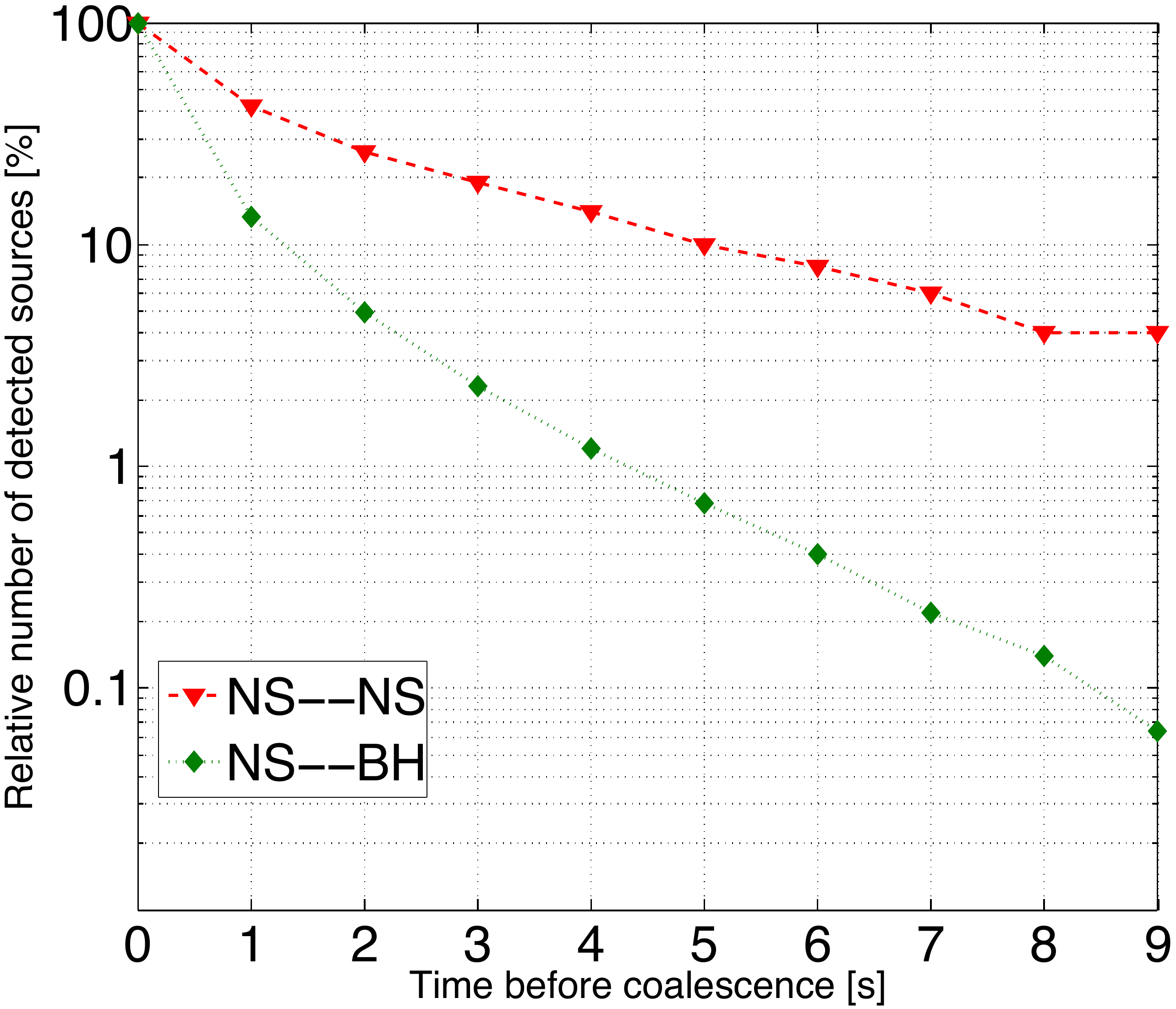}
\caption{Number of detectable NS--NS signals (red) and NS--BH signals (green) as a function of time before coalescence.}
\label{comparedistro}
\end{center}
\end{figure}

There were several analytical \cite{Fairhurst:2009tc,Fairhurst:2010is} and numerical \cite{Nissanke:2011ax} attempts to obtain a relation between time accuracy and the sky area. 
We will follow the analytic way laid out in Ref.~\cite{Fairhurst:2010is} to obtain an approximate area given the observation of a binary signal. 
We approximate the area at a given confidence level CL as 

\begin{equation}
\label{eq:areatheta}
A(CL)\simeq2\pi\sigma_x\sigma_y\left[ -\ln(1-CL) \right]/\cos{\theta}\;,
\end{equation}

\noindent
where $\sigma_x$ and $\sigma_y$ denote the errors associated with the eigen-directions of the matrix describing the location accuracy (see Ref.~\cite{Fairhurst:2009tc,Fairhurst:2010is} for details), and $\theta$ is the angle between the normal of the plane spanned by the three detectors and the direction to the source. 

Since the two LIGO detectors have comparable sensitivity and the light-travel time from each LIGO site to Virgo is about the same (27 ms), it is possible to simplify the analytical sky location expression of Ref.~\cite{Nissanke:2011ax} by taking the timing accuracy $\sigma_{l}$ of the two LIGO detectors to be identical, but allowing the Virgo timing $\sigma_{v}$ to differ.
Using a median angle as  $\cos{\theta}\simeq 0.5$, the $90\%$ confidence level sky area $A$ is 

\begin{equation}
A_\mathrm{signal} \simeq 20  \mathrm{deg}^{2} \left(\frac{\sigma_{l}}{0.25 \mathrm{ms}}\right) \left(\frac{\sqrt{(2\sigma_{v}^{2}+\sigma_{l}^{2})/3)}}{0.25 \mathrm{ms}}\right)\;.
\label{eq:approxarea}
\end{equation}

Note, that with a network composed of three detectors, the sky area is degenerate since such a network cannot discriminate between a location above or below the plane spanned by the detectors.

Since the error on the arrival time of a signal can be determined for each detector by using the bandwidth, it is important to note the properties of bandwidth before the coalescence. 
The dependence of the bandwidth and the SNR prior to coalescence is illustrated in Fig.~\ref{fig:snr_band}. Although the SNR rises reasonably fast before the time of the merger, the bandwidth stays at very small values even very close to the merger because of the steep rise of the seismic wall in the power spectrum of the advanced detectors at lower frequencies (see Fig. \ref{powers}).  

\subsubsection{Localization with four or more detectors}
\label{sec:loc4}

The approximation described in the previous section to obtain the sky area in the case of a three detector network cannot be applied to networks with four or more detectors.  Thus a different approach must be taken when dealing with a network with more detectors.
In that approach, the whole sky is tiled into equidistant sky points, and the expected time delay between each detectors $t_{AB}^P$ is calculated for each point $P$. The time residual for each detector combination is 

\begin{equation}
 \frac{t_{AB} - t_{AB}^P}{\Delta t_{AB}}\;,
\end{equation}

\noindent
where $t_{AB}$ is the observed light-travel time and $\Delta t_{AB}$ is the error on the expected timing accuracy given as $ \Delta t_{AB} = \sqrt{\sigma_{t,A}^2 + \sigma_{t,B}^2} $ for detectors $A$ and $B$.
By computing the residual sum of squares over each detector combination \textit{combo},

\begin{equation}
\Delta T^2_{rss} = N\sum_{\mathrm{combo}} \frac{t_{AB} - t_{AB}^P}{\Delta t_{AB}}\;,
\end{equation}

\noindent
a probability can be assigned to each sky point $P$:

\begin{equation}
p \propto \frac{1}{2}e^{-\Delta T_{rss}^2/2}\;.
\label{eq:skyarea}
\end{equation}

\noindent
After normalization of this quantities, all sky areas corresponding to the sky points are summed until the desired confidence level is reached. 
The resulting area $A_\mathrm{signal}$ is the sky area where the source is located at the given confidence level. This method is described in more detail in Ref.~\cite{DietzSkyLoc}.

\section{Results}
\label{sec:results}

\subsection{Single detector results}

Fig.~\ref{fig:snr_band} shows the evolution of the accumulated SNR and frequency bandwidth over the course of the inspiral for the NS--NS and the NS--BH case.
In the case of a NS--BH binary, the SNR is accumulated much closer to coalescence compared to a NS--NS binary  (see Table~I). 
Since the gravitational wave signal from a NS--BH binary has smaller frequencies close to coalescence than a NS--NS binary, these systems reach the sensitive spectral part of a detector later, resulting in slower accumulation of  SNR.
Another interesting fact is the moment in which s signal enters the detector band. A signal from a NS--NS binary enters the sensitive band about 52 seconds prior coalescence, while it is only 12 seconds for a NS--BH system, given a frequency threshold of 30~Hz. 
If the frequency threshold is at 10~Hz, the NS--BH enters the detector band 215 seconds before coalescence. Therefore, in order to detect more massive systems, a lower frequency threshold becomes crucial. 
Fig.~\ref{comparedistro} shows the fraction of detectable inspiral systems as time prior to coalescence, compared to the numbers detected at the time of coalescence.
The fraction of NS--NS systems is much higher than that of NS--BH systems. At a time $\sim$10 seconds prior to coalescence about 5\% of the NS--NS systems are detectable, but only about 0.5\% of the NS--BH systems.

\begin{table}[b!]
\caption{\label{table:cumultable} Fraction of the SNR as function of the time before coalescence for NS--NS and NS--BH systems, compared to the SNR accumulated at the time of coalescence.}
\begin{indented}
\item[]\begin{tabular}{cccccc}
\hline
   & 10\% & 25\% & 50\% & 75\% & 90\% \\
   \hline
  NS--NS & 104s & 21s & 4.8s & 1s & 0.3s \\
  \hline
  NS--BH & 23s & 4.7s & 1.1 s & 0.23s & 0.06s\\
  \hline 

\end{tabular}
\end{indented}
\end{table}

\begin{figure}[t!]
\begin{center}
\includegraphics[scale=0.3]{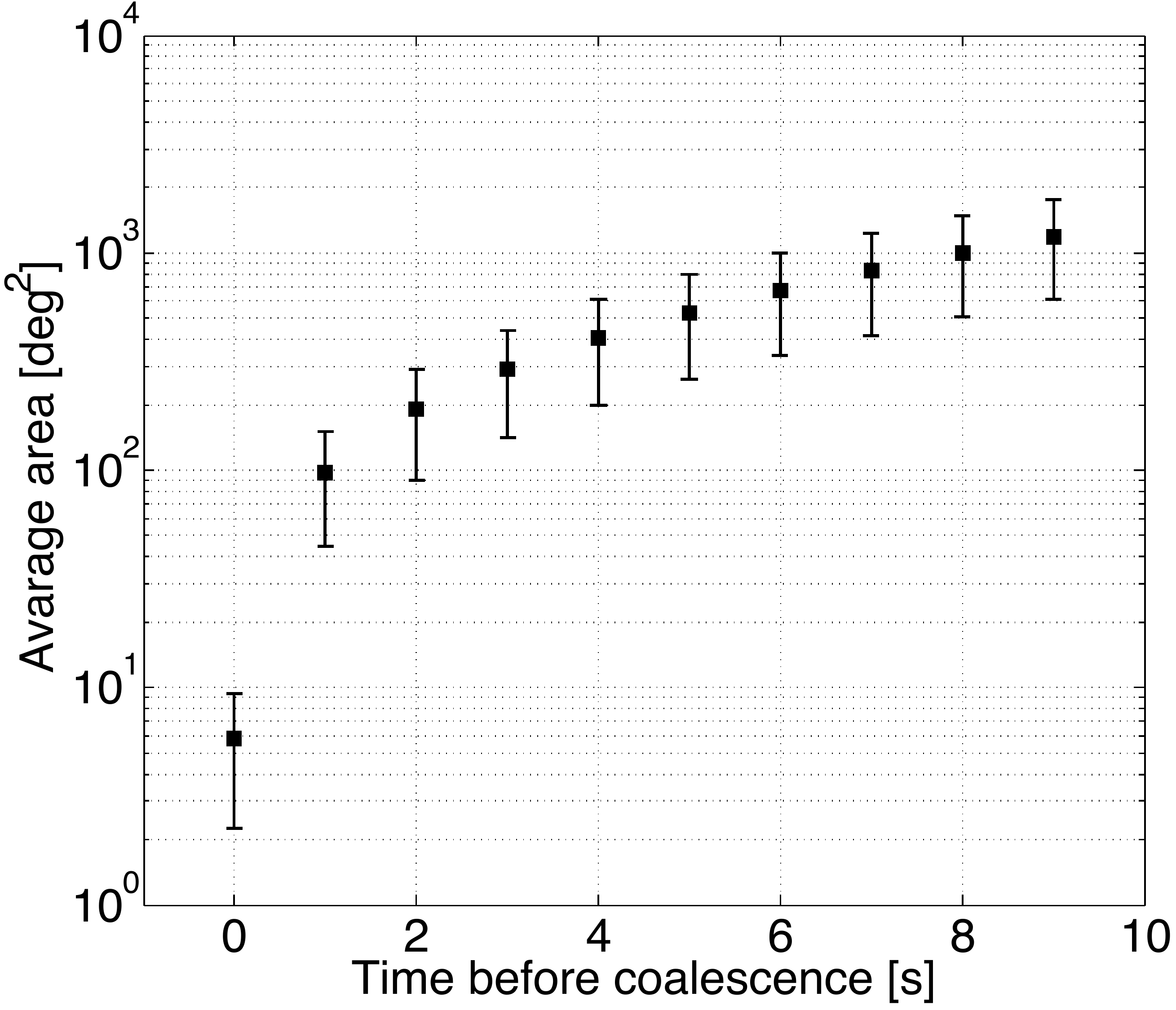}
\caption{Average sky area for a NS--NS system as a function of time. The error bars denote the standard deviation from the simulated sample.}
\label{fig:meanarea}

\end{center}
\end{figure}

Early detection of a signal is only the first step to generate a notification for a telescope. 
The second step, the localization of the source, strongly depends on the timing accuracy of the signal.
Fig.~\ref{fig:snr_band} shows that the bandwidth, and consequently the timing accuracy via Eq.~(\ref{accuracy}), is largely accrued in the last moments prior to coalescence. 
In the next section we show that detecting and localizing a signal with a three-detector network alone is very challenging. 
Adding a fourth detector, like the Japanese KAGRA detector, results in dramatically increased detectability and location accuracy.

\subsection{Network Results}

In this section, the Monte Carlo simulation is applied to a detector network.
A source is defined as detected, when the SNR in each detector $i$ is $\rho_{i}>7$. This choice of threshold has been used in previous searches \cite{Aylott:2009ya}.

Future large-scale transient survey projects like the Large Synoptic Survey Telescope (LSST)~\cite{urlLSST}, have an expected  field-of-view (FOV) of $\sim$10 square degrees. 
We estimate the probability $p$ of observing the source of a GW signal inside that FOV by dividing the FOV by our estimated sky area. If the estimated sky area is smaller than 10 square degrees, the probability is set to 1:

\begin{equation}
 p =\left\{
  \begin{array}{l l}
   10\, \mathrm{deg}^2/A_\mathrm{signal}& \mathrm{if } A_\mathrm{signal} \geq 10\, \mathrm{deg}^2 \\
   1                                  & \mathrm{if } A_\mathrm{signal}  <10\, \mathrm{deg}^2 \;.
  \end{array}\right.
\end{equation}

\begin{figure}[t!]
\begin{center}
\includegraphics[scale=0.4]{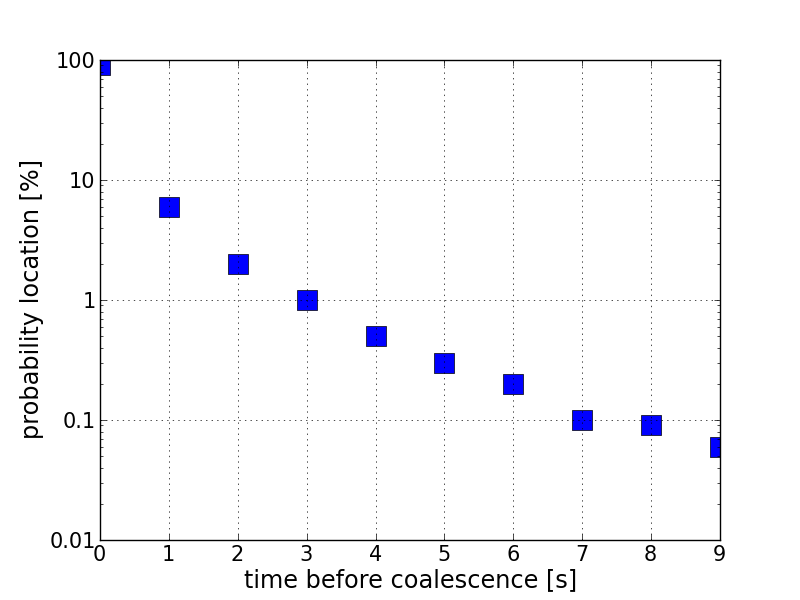}
\caption{Probability of a GW-optical detection for a NS--NS system as a function of time before coalescence for the three detector network. }
\label{fig:probability}
\end{center}
\end{figure}

For the three-detector LV network, the probabilities are listed in Table~II and plotted in Fig.~\ref{fig:probability}.
The approximate average area for a signal detected prior to coalescence is shown in Fig.~\ref{fig:meanarea}.
 Although the probability for catching the correct location at the time of coalescence is 90\%, the probability decreases very fast and gets well below 1\% for a signal detected $\overset{\sim}{<}$4 seconds prior to coalescence.

Simulating the four-detector LVK network, the outcome is quite different. 
The additional detector in Japan introduces a network baseline, which is larger than for any detector pair  of the LV network. 
Furthermore, the addition of a fourth detector eliminates the ambiguity of the sky area.
However, because the approximation Eq.~(\ref{eq:approxarea}) can only be applied to a three detector network, we need to use the formalism for the general case as described in section~\ref{sec:loc4}. 

When performing the Monte Carlo simulation with the LVK network, we found very encouraging results. 
Even for the most difficult case considered in this work,  the detection and location of a NS--NS signal $\sim$10 seconds prior to coalescence, the sky area is determined very precise. 
In 98.5\% of all cases, the sky area of the detected sources were localized to a sky area smaller than 10 square-degree. 
Since about 50 NS--NS sources a year are expected to be observed by advanced detectors~\cite{LIGO:2010cfa}, about 2-3 sources may be detected and localized at $\sim$10 seconds prior to coalescence.
This opens very exciting prospects for science as the coalescence of a merger may be observed in coincident by GW detectors and optical telescopes! 
In case of NS--BH systems those prospects are less encouraging, since the fraction that may be detected that early in a GW detector is only about 0.5\%. 
However, the rate expectations for those systems are more uncertain, so it is very hard to make an adequate prediction. 

In order to get the most science of a coincident observation, an electromagnetic telescope must be pointed in the right direction in time. 
The whole data processing, data transfer and telescope pointing must not take more than 10 seconds.
First, the data itself must be recorded at the detector sites, analyzed, and transferred to a central place, where the coincidence step is performed. This step might be accomplished within about 2 seconds~\cite{Cannon:2011vi}.
Second, if the analysis detects a coincident signal between the detectors, the sky area need to be computed and transferred to participating telescopes. This step may also be done within about 2 seconds. 
The third and last step, the actual telescope pointing, is most time consuming. For example, the pointing time of the  Robotic Optical Transient Search Experiment (ROTSE) \cite{Akerlof:2002xu} is about 4~seconds.
To conclude, the total process from detecting a signal to the time a telescope is on target may be done in just under 10 seconds. 

\begin{figure}[t!]
\begin{center}
\includegraphics[scale=0.45]{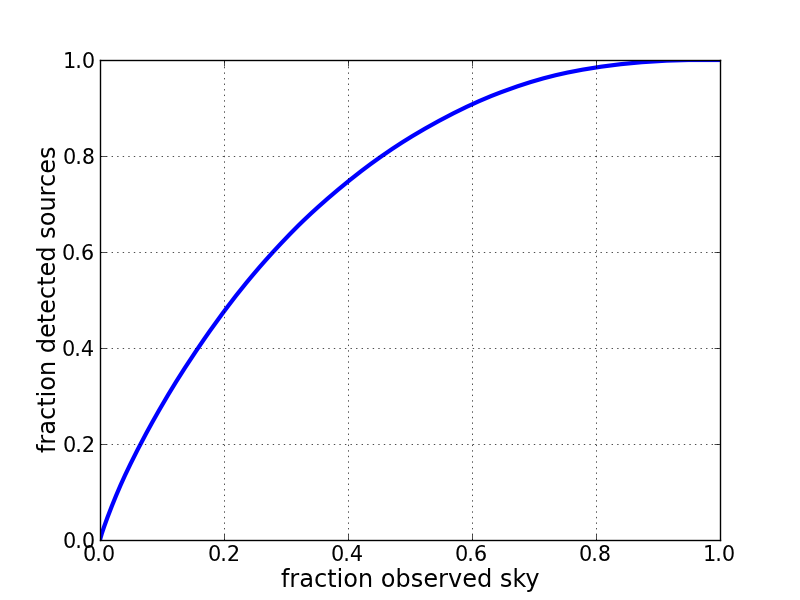}
\caption{Fraction of detected sources as a function of the observed fraction of the sky. As the regions of the sky are known, in which the detector network can see farthest, a array of telescopes only need to cover some fraction of the sky. Covering 40\% of the sky yield in a probability of around 80\% to have the source in the telescope's FOV.}
\label{fig:fractionSkyDetect}
\end{center}
\end{figure}

The sensitivity of a GW network is not uniform over the sky. 
The network is more sensitive to some sky regions than others, which can be seen in the sky maps shown in Fig.~\ref{fig:response-network} and \ref{fig:response-network4}. 
The more sensitive regions on the sky are known a priori, and it is more likely a source if detected in this region. 
Telescopes could be pointed in such regions beforehand, or even an array of fixed mounted telescopes can be used to observe the interesting regions of the sky. 
 Fig.~\ref{fig:fractionSkyDetect} shows the fraction of the sky that need to be observed in order to make a detection of a signal at given probability. 
For example, if the GW network detects a signal from a GW source, a telescope array which covers 40\% of the whole sky should be able to catch this signal with a chance of $\sim$80\% .

\begin{table}[b!]
\caption{\label{tab:prob3} Probabilities to have a 10-deg$^2$ telescope pointed to the correct location, in case the data analysis, the flow of information and the telescope's pointing time are negligible.}
\begin{indented}
\item[] \begin{tabular}{cc}
\hline
   time [s] & probability \\
   \hline
      0 & 90\% \\
   \hline
 1 & 6\% \\
   \hline
   2 & 2\% \\
   \hline
   3 & 1\% \\
   \hline
   4 & 0.5\% \\
   \hline
   5 & 0.3\% \\
   \hline
   6 & 0.2\% \\
   \hline
   7 & 0.1\% \\
   \hline
   8 & 0.09\% \\
   \hline
   9 & 0.06\% \\
   \hline
  
\end{tabular}
\end{indented}
\end{table}

\section{Discussion}
\label{sec:conclusion}

The work presented above shows that inspiral signals from the merger of two compact objects can be detected and localized \textit{before} its coalescence. 
Although the majority of SNR is accumulated in the last moments before coalescence, an early detection and localization could become very likely with advanced detectors.
Two GW detector networks have been considered, one consisting of the two aLIGO and the aVirgo detectors, and one network with the Japanese KAGRA in addition. 
The approximate SNR of simulated source is calculated using the expected spectral density noise curves.  
A source is considered detected, if the  SNR is above 7 in each detector.
The number of detectable sources decrease very steep at times prior to coalescence.
Only $\sim$3\% of NS--NS are detectable at $\sim$10 seconds prior to coalescence, while this number becomes  $\sim$0.6\% for NS--BH sources. 
As the expected number of coalescences detectable with advanced LIGO/Virgo is around 50 per year for NS--NS binaries, it might be possible to detect a few of such sources at about 10 seconds before the coalescence. 

Having made a detection, the sky area must be computed from the observed signal.
If the signal is observed with the 3-detector network, the sky areas are typically hundreds of square-degrees, which makes it very unlikely to have an array of telescopes pointed in time to the correct direction. 

However, when including KAGRA the prospects of early detection and location become much more likely. As KAGRA introduces a very long additional baseline to the network, avoids the ambiguity of a 3-detector network, and provides additional information to determine the sky area, it seems possible a single 10 square-degree telescope is enough to observe the sky area around the coalescence.

The question arises if there is enough time to process the GW data, and to send a notification to participating telescopes.  With data-processing tools currently developed for advanced LIGO/Virgo, the data-analysis time can be reduced to virtually zero time.
The time required to send the data between the detector sites and to the telescopes can also assumed to be sufficiently small. 
The most time consuming step seem to be the telescope pointing itself, but current telescopes are able to accomplish the pointing within a few seconds. 
Finally, optical observations during the coalescence of two compact objects seems possible. 

Even if a coincident GW/optical observation is possible only in a very few cases per year, this opens incredible scientific possibilities.
It may become possible to directly test the connection of binary mergers with gamma-ray bursts, and to constrain models of their central engines. 
GRB outflow models can be tested, the geometry of the outflow, and tests of Lorentz invariance and tests of General Relativity in the strong field regime become possible. 
A complete overview of the implications of such a coincident observation is beyond the scope of this work, and subject of another investigation.

\ack
The work reported here is the result a summer student project at the University of Mississippi. AM and his supervisor, AD, were supported by the National Science Foundation through awards PHY-0757937 and PHY-1067985. 
The authors also would like to thank Marco Cavaglia for his support and suggestions regarding this work.

\end{document}